
\newif\iffigs\figstrue

%
\let\useblackboard=\iftrue
%
%
\newfam\black

\input harvmac.tex

\iffigs
  \input epsf
\else
  \message{No figures will be included.  See TeX file for more
information.}
\fi

\def\Title#1#2{\rightline{#1}
\ifx\answ\bigans\nopagenumbers\pageno0\vskip1in%
\baselineskip 15pt plus 1pt minus 1pt
\else
\def\listrefs{\footatend\vskip 1in\immediate\closeout\rfile\writestoppt
\baselineskip=14pt\centerline{{\bf References}}\bigskip{\frenchspacing%
\parindent=20pt\escapechar=` \input
refs.tmp\vfill\eject}\nonfrenchspacing}
\pageno1\vskip.8in\fi \centerline{\titlefont #2}\vskip .5in}

\ifx\answ\bigans\def\tcbreak#1{}\else\def\tcbreak#1{\cr&{#1}}\fi
\useblackboard
\message{If you do not have msbm (blackboard bold) fonts,}
\message{change the option at the top of the tex file.}
\font\blackboard=msbm10 
\font\blackboards=msbm7
\font\blackboardss=msbm5
\textfont\black=\blackboard
\scriptfont\black=\blackboards
\scriptscriptfont\black=\blackboardss
\def\Bbb#1{{\fam\black\relax#1}}
\else
\def\Bbb#1{{\bf #1}}
\fi
%
\def\yboxit#1#2{\vbox{\hrule height #1 \hbox{\vrule width #1
\vbox{#2}\vrule width #1 }\hrule height #1 }}
\def\fillbox#1{\hbox to #1{\vbox to #1{\vfil}\hfil}}
\def\ybox{{\lower 1.3pt \yboxit{0.4pt}{\fillbox{8pt}}\hskip-0.2pt}}
\def\np#1#2#3{Nucl. Phys. {\bf B#1} (#2) #3}

\def\comments#1{}

\def\CA{{\cal A}}

\def\CF{{\cal F}}

\def\II{\relax{I\kern-.07em I}}

\def\IZ{\relax\ifmmode\mathchoice
{\hbox{\cmss Z\kern-.4em Z}}{\hbox{\cmss Z\kern-.4em Z}}
{\lower.9pt\hbox{\cmsss Z\kern-.4em Z}}
{\lower1.2pt\hbox{\cmsss Z\kern-.4em Z}}\else{\cmss Z\kern-.4em
Z}\fi}
\def\IB{\relax{\rm I\kern-.18em B}}
\def\IC{{\relax\hbox{$\inbar\kern-.3em{\rm C}$}}}
\def\ID{\relax{\rm I\kern-.18em D}}
\def\IE{\relax{\rm I\kern-.18em E}}
\def\IF{\relax{\rm I\kern-.18em F}}
\def\IG{\relax\hbox{$\inbar\kern-.3em{\rm G}$}}
\def\IGa{\relax\hbox{${\rm I}\kern-.18em\Gamma$}}
\def\IH{\relax{\rm I\kern-.18em H}}
\def\II{\relax{\rm I\kern-.18em I}}
\def\IK{\relax{\rm I\kern-.18em K}}
\def\IP{\relax{\rm I\kern-.18em P}}

\useblackboard
\def\IZ{\relax\Bbb{Z}}
\fi

\font\cmss=cmss10 \font\cmsss=cmss10 at 7pt
\def\IR{\relax{\rm I\kern-.18em R}}

\def\bR{\bf R}
\def\bS{\bf S}
\def\tilde{\widetilde}
\def\Def{\mathop{\rm Def}\nolimits}

\Title{ \vbox{\baselineskip12pt\hbox{hep-th/9609070, DUKE-TH-96-130,
\hbox{RU-96-80}}}}
{\vbox{
\centerline{Extremal Transitions and Five-Dimensional}
\vskip2pt\centerline{Supersymmetric Field Theories}}}
\centerline{David R. Morrison}
\smallskip
\smallskip
\centerline{Department of Mathematics, Box 90320}
\centerline{Duke University}
\centerline{Durham, NC 27708-0320 USA}
\centerline{\tt drm@math.duke.edu}
\bigskip
\centerline{and}
\bigskip
\centerline{Nathan Seiberg}
\smallskip
\smallskip
\centerline{Department of Physics and Astronomy}
\centerline{Rutgers University }
\centerline{Piscataway, NJ 08855-0849}
\centerline{\tt seiberg@physics.rutgers.edu}
\bigskip
\bigskip
\noindent
We study five-dimensional supersymmetric field theories with one-dimensional
Coulomb branch.  We extend a previous analysis which led to non-trivial fixed
points with $E_n$ symmetry ($E_8$, $E_7$, $E_6$, $E_5=Spin(10)$, $E_4=SU(5)$,
$E_3=SU(3)\times SU(2)$, $E_2=SU(2)\times U(1)$ and $E_1=SU(2)$) by finding two
new theories: $\tilde E_1$ with $U(1)$ symmetry and $E_0$ with no symmetry.
The latter is a non-trivial theory with no relevant operators preserving the
super-Poincar\'e symmetry.  In terms of string theory these new field theories
enable us to describe compactifications of the type I$'$ theory on
$\bS^1/\IZ_2$
with 16, 17 or 18 background D8-branes. These theories also play a crucial role
in compactifications of M-theory on Calabi--Yau spaces, providing physical
models for the contractions of del Pezzo surfaces to points (thereby completing
the classification of  singularities which can occur at codimension one in
K\"ahler moduli).  The structure of the Higgs branch
yields a prediction which unifies the known mathematical facts about del Pezzo
transitions in a quite remarkable way.

\Date{September, 1996}

\lref\vafa{C. Vafa, Private communication.}

\lref\dkv{M. Douglas, S. Katz, and C. Vafa, to appear.}

\lref\MVii{D. R. Morrison and C. Vafa, ``Compactifications of F-Theory on
Calabi--Yau Threefolds (II),'' \np{476}{1996}{437--469}, {\tt
hep-th/9603161}.}

\lref\manin{Yu. I. Manin, {\it Cubic Forms: Algebra, Geometry, Arithmetic},
2nd ed., North-Holland, Amsterdam, New York, 1986, Chapter IV.}

\lref\demazure{M. Demazure, ``Surfaces de del Pezzo, II, III, IV, V,''
S\'eminaire sur les
Singularit\'es des Surfaces, Lecture Notes in Math. vol. 777,
Springer-Verlag, 1980, pp.~21--69.}

\lref\look{D. R. Morrison, ``Through the Looking Glass,'' to appear.}

\lref\mori{S. Mori, ``Threefolds Whose Canonical Bundles are Not
Numerically Effective,'' Annals of Math. (2) {\bf 116} (1982) 133--176.}

\lref\friedman{R. Friedman,
``Simultaneous Resolution of Threefold Double Points,''
Math. Ann. {\bf 274} (1986) 671--689.}

\lref\clemens{H. Clemens,
``Double Solids,''  Adv. Math. {\bf 47} (1983) 107--230.}

\lref\schoen{C. Schoen,  ``On Fiber Products of Rational Elliptic
Surfaces with Section,'' Math. Zeit.  {\bf 197} (1988) 177--199.}

\lref\hayakawa{Y. Hayakawa, ``Degeneration of Calabi--Yau Manifold with
Weil--Petersson Metric,'' {\tt alg-geom/9507016}.}

\lref\reidcanon{M. Reid, ``Canonical 3-Folds,'' Journ\'ees de
G\'eom\'etrie Alg\'ebriqe d'Angers (A. Beauville, ed.),
Sitjhoff \& Noordhoof,  1980, pp.~273--310.}

\lref\reiddelP{M. Reid, ``Nonnormal del Pezzo Surfaces,''
Publ. Res. Inst. Math. Sci. {\bf 30} (1994) 695--727,
{\tt alg-geom/9404002}.}

\lref\grossobstr{M. Gross, ``The Deformation Space of Calabi--Yau
$n$-Folds Can Be Obstructed,'' Mirror Symmetry II (B. Greene and S.-T.
Yau, eds.), to appear, {\tt alg-geom/9402014}.}

\lref\grossdef{M. Gross, ``Deforming Calabi--Yau Threefolds,''
{\tt alg-geom/9506022}.}

\lref\grossprim{M. Gross, ``Primitive Calabi--Yau Threefolds,''
{\tt alg-geom/9512002}.}

\lref\namikawa{Yo. Namikawa, ``Stratified Local Moduli of Calabi--Yau
3-Folds,'' preprint, 1995.}

\lref\wilson{P. M. H. Wilson, ``The K\"ahler Cone on Calabi--Yau
Threefolds,'' Invent. Math. {\bf 107} (1992) 561--583; Erratum,
ibid. {\bf 114} (1993) 231--233.}

\lref\quotient{M. Schlessinger, ``Rigidity of Quotient Singularities,''
Invent. Math. {\bf 14} (1971) 17--26.}

\lref\kleplak{H. Kleppe and D. Laksov, ``The Algebraic Structure and
Deformation of Pfaffian Schemes,'' J. Algebra {\bf 64} (1980) 167--189.}

\lref\altmann{K. Altmann, ``The Versal Deformation of an Isolated Toric
Gorenstein Singularity,'' {\tt alg-geom/9403004}.}

\lref\BE{D. Buchsbaum and D. Eisenbud, ``Algebra Structures for Finite
Free Resolutions and Some Structure Theorems for Ideals of Codimension
3,'' Amer. J. Math. {\bf 99} (1977) 447--485.}

\lref\rStrominger{A. Strominger, ``Massless Black Holes and Conifolds in
String Theory,'' Nucl. Phys. {\bf B451} (1995) 97--109, {\tt
hep-th/9504090}.}

\lref\rGMS{B. R. Greene, D. R. Morrison, and A. Strominger,
     ``Black Hole Condensation and the Unification of String Vacua,''
     Nucl. Phys. {\bf B451} (1995) 109--120, {\tt hep-th/9504145}.}

\lref\rGMV{B. R. Greene, D. R. Morrison, and C. Vafa,
``A Geometric Realization of Confinement,'' {\tt hep-th/9608039}.}

\lref\rKMP{S.~Katz, D.~R. Morrison, and
M.~R. Plesser, ``Enhanced Gauge Symmetry in Type II String Theory,''
{\tt hep-th/9601108}.}

\lref\klkam{P. Berglund, S. Katz,
A. Klemm, and P. Mayr, ``New Higgs Transitions Between Dual N=2 String
Models,'' {\tt hep-th/9605154}.}

\lref\fived{N. Seiberg, ``Five Dimensional SUSY Field Theories,
Non-trivial Fixed Points, and String Dynamics,'' {\tt hep-th/9608111}.}

\lref\ccdf{A. C. Cadavid, A. Ceresole, R. D'Auria, and S. Ferrara,
``Eleven-Dimensional Supergravity Compactified on a Calabi--Yau
Threefold,''
Phys. Lett. {\bf B357} (1995) 76--80, {\tt hep-th/9506144}.}

\lref\fkm{S. Ferrara,
R.R. Khuri, and R. Minasian, ``M-Theory on a Calabi--Yau Manifold,''
Phys.Lett. {\bf B375} (1996) 81--88  {\tt hep-th/9602102}.}

\lref\fms{S. Ferrara,
R. Minasian, and A. Sagnotti, ``Low Energy Analysis of $M$ and $F$
Theories on Calabi--Yau Threefolds'', {\tt hep-th/9604097}.}

\lref\mandf{E. Witten, ``Phase
Transitions in M-Theory and F-Theory,'' {\tt hep-th/9603150}.}

\lref\nclp{For a nice review see, S. Chaudhuri, C. Johnson, and J.
Polchinski, ``Notes on D-Branes,'' {\tt hep-th/9602052}.}

\lref\threedone{N. Seiberg, ``IR Dynamics on Branes and Space-Time
Geometry,'' {\tt hep-th/9606017}.}

\lref\nonren{N. Seiberg, ``Naturalness Versus Supersymmetric
Non-renormalization Theorems,'' Phys. Lett. {\bf B318} (1993) 469--475,
{\tt hep-ph/9309335}.}

\lref\aps{P.~C. Argyres, M.~R. Plesser, and N. Seiberg,
``The Moduli Space of Vacua of $N=2$ SUSY QCD and Duality in $N=1$ SUSY
QCD,'' \np{471}{1996}{159--194}, {\tt hep-th/9603042}.}

\lref\polwit{J. Polchinski and E. Witten, `` Evidence for Heterotic --
Type I Duality,'' \np{460}{1996}{525--548}, {\tt hep-th/9510169}.}

\lref\rBBS{K. Becker, M. Becker, and A. Strominger,
     ``Fivebranes, Membranes and Non-Per\-tur\-ba\-tive String Theory,''
     Nucl. Phys. {\bf B456} (1995) 130--152,
     {\tt hep-th/9507158}.}

\newsec{Introduction}

Local quantum field theories occur in string theory in two places.  They
describe both the low energy excitations in string compactifications and
those on D-branes \nclp.  Therefore, we have three distinct parallel
lines of research: local quantum field theory, the theory of D-branes,
and string compactification.  It is extremely interesting to understand
the relation between these three topics and to establish a complete
dictionary translating concepts {}from one topic to the others.

The traditional approach has been to use field theory results to learn
about properties of string theory.  Recently, however, the opposite
logic turned out to be fruitful.  The assumption of string duality was
used to shed new light on interesting field theories
\refs{\threedone, \fived}.  In particular, in \fived\ the
five-dimensional field theory on a D4-brane probe was studied.  The
analysis led to the discovery of new non-trivial fixed points.  As we
said above, these theories also appear in string compactifications.
Here we complete the picture by relating this new field theoretic
understanding to the geometry of the compactification.

In section 2 we will review the analysis of \fived\ and will extend it.
We will consider $U(1)$ and $SU(2)$ field theories with $N_f$ quark
flavors and will describe their low energy field theory.  We will show
how non-trivial fixed points with $E_n$ symmetry ($E_8$, $E_7$, $E_6$,
$E_5=Spin(10)$, $E_4=SU(5)$, $E_3=SU(3)\times SU(2)$, $E_2=SU(2)\times
U(1)$ and $E_1=SU(2)$) can be found.  We will also show that upon
suitable perturbations these theories can flow to two new theories:
$\tilde E_1$ with $U(1)$ symmetry and $E_0$ with no symmetry.

These results are derived using the assumption of type I/type
I$'$/heterotic duality.  They enable us to extend the type I$'$
description to the entire moduli space of vacua including vacua where at
some point in space-time the coupling constant diverges \refs{\polwit,
\nclp}.  We find that while in type I$'$ compactifications on
$\bS^1/\IZ_2$ at weak coupling there are two orientifolds and 16
D8-branes, at strong coupling there can be 17 or 18 D8-branes.  This can
happen only when the string coupling at one or at both ends of
$\bS^1/\IZ_2$ diverges.

The other application of these field theories is discussed in section 3.
Five-dimensional field theories occur in compactifications of M-theory
on a Calabi--Yau threefold.  Such compactifications were studied in
\refs{\ccdf\fkm\mandf{--}\fms}.  Of particular interest are the
compactifications
on singular Calabi--Yau spaces.  Following Strominger \rStrominger, we
should understand these singularities in terms of new degrees of freedom
which become massless at that point.  Such degrees of freedom can enable
one to understand transitions to different compactifications \rGMS.
Witten has performed such an analysis in \mandf\ for two kinds of
singularities.  As in four dimensions, the conifold singularity
\refs{\rStrominger, \rGMS} was interpreted as an Abelian gauge theory
with new massless hypermultiplets and the extremal transition of
\refs{\rKMP,\klkam} was interpreted as an enhanced non-Abelian gauge
theory.

Here, we complete the story by identifying the physics of the third and
final type of singularity whose geometry is that of a four-manifold
within the Calabi--Yau (a so-called ``del Pezzo surface'') which shrinks
to zero size at the transition point. These turn out to be the
interacting supersymmetric five-dimensional theories mentioned
above.\foot{The observation that the $E_n$ theories of \fived\ explain
the physics of del Pezzo contractions was independently made by Vafa
\vafa, and is being further developed by Douglas, Katz, and Vafa \dkv.}
In such interacting conformal field theories the notion of a particle is
ill defined.  Therefore, it is meaningless to ask which particles are
massless at that point in the moduli space.

The known mathematical classification of del Pezzo surfaces precisely
matches the list of interacting field theories we have found.  Furthermore,
the identification of the symmetry group at the transition point as
one of the $E_n$ groups yields the prediction that the Higgs branch of these
theories will coincide with the corresponding instanton moduli space.  That
prediction
unifies the known mathematical facts about these del Pezzo transitions
in a quite remarkable way, and suggests a uniform structure which had
not previously been suspected.

\newsec{Five-dimensional supersymmetric field theories}

In this section we review and extend the discussion in \fived\ of
five-dimensional supersymmetric field theories.  The massless fields are
in hypermultiplets or vector multiplets.  These theories typically have
a moduli space of vacua.  The scalars in the hypermultiplets
parameterize a hyper-K\"ahler manifold which we will refer to as the
Higgs branch.  The scalars in the vector multiplets $\CA^i$ parameterize
the Coulomb branch.  The metric on the Coulomb branch is determined by a
prepotential $\CF(\CA^i)$ which is locally cubic
\eqn\metric{(ds)^2 = {\partial^2 \CF \over \partial \CA^i \partial
\CA^j} d\CA^id\CA^j.}
The metric determines the kinetic terms of the scalars in the multiplets
and the gauge coupling constants for the ``photons'' in the vector
multiplets.  Since $\CF$ is cubic, the metric varies linearly.  The
singularities in the Coulomb branch are of two kinds.  First, the moduli
space of vacua could have boundaries.  Second, the space could be smooth
but the metric \metric\ could be non-differentiable.  At such points the
slope of the metric changes.

As in \fived, we limit ourselves to theories with a one-dimensional
Coulomb branch parameterized by a real scalar $\phi$.  The
metric and the gauge coupling are determined by a piecewise linear
function
\eqn\deft{t(\phi)= {1 \over g^2(\phi)}= t_0+c\phi.}
Singularities of the first kind correspond to a
Coulomb branch being $\bR^+$.  The gauge coupling near the origin is of
the form \deft.  If $c>0$, we could also have $t_0=0$.
Singularities of the second kind have a Coulomb branch $\bR$.  Around
the singular point $\phi_0$ the metric is determined by
\eqn\defts{t(\phi)= t_0+c_0\phi + c |\phi -\phi_0|.}
The critical behavior (the singularity) is again characterized by the
value of $c$.

The simplest theories studied in \fived, are $U(1)$ gauge theories with
$N_f$ ``electrons.''  At tree level $t=t_0$ is independent of $\phi$.
However, a one-loop correction \mandf\ induces non-zero $c$ such that
\eqn\tuone{t(\phi)=t_0 - \sum_{i=1}^{N_f} |\phi - m_i|}
where $m_i$ are the electron masses.  We would like to make a few
comments on \tuone\

\item{(1.)} By shifting $\phi$ we can change $m_i$ such that only
$N_f-1$ of them are important.

\item{(2.)} The one-loop computation is divergent.  Therefore, we had to
absorb an additive (non-universal) divergent renormalization in $t_0$.

\item{(3.)} There could also be a finite correction to $t_0$.  On
dimensional grounds, any such correction must be
linear in the masses $m_i$.  Using the
symmetries, it must also be proportional to $\sum_i m_i$.  We can absorb
such a potential contribution in $t_0$.

\item{(4.)} The expression \tuone\ makes sense only when $t(\phi)$ is
non-negative.  This means that $t_0$ should be positive and the $m_i$ are
constrained.  Furthermore, for any $t_0$ and $m_i$, $\phi$ is bounded.
In other words, the moduli space always has singularities.  These
reflect the lack of renormalizability of the theory, signaling that it
must be embedded in another theory at short distance.

The gauge coupling $g={1 \over \sqrt t}$ has negative mass dimension and
therefore it is ``irrelevant'' at long distance.  Hence, the degrees of
freedom at the singularities are the fundamental photon and electrons.

When several masses are equal, several singularities in \tuone\
coalesce.  Consider the case where all the masses are equal.  Shifting
$\phi$ we can set the mass to zero.  The global symmetry of the theory
is then $SU(N_f)=A_{N_f-1}$ (it is not $U(N_f)$ because the $U(1)$
factor is gauged).  Equation \tuone\ becomes
\eqn\tuonenf{t(\phi)=t_0 - N_f |\phi|.}
The degrees of freedom at the singularity are again the elementary
photon and electrons.  There is a Higgs branch emanating {}from the
origin.  It is isomorphic to the moduli space of $SU(N_f)$ instantons.

We will refer to these theories as $A_{N_f-1}$ theories reflecting their
global symmetry and their Higgs branch.  We will extend this terminology
to the case of $N_f=1$ and will refer to this latter theory as $A_0$.

The second class of theories consists of $SU(2)$ gauge theories with $N_f$
``quarks'' which are hypermultiplets in the two-dimensional
representation of the gauge group.  Now the Coulomb branch is $\bR^+$
parameterized by $\phi \ge 0$.  The one-loop computation yields
\eqn\tsutwo{t(\phi)=t_0 +16\phi - \sum_{i=1}^{N_f} |\phi - m_i| -
\sum_{i=1}^{N_f} |\phi + m_i|.}
We will make similar  comments to the ones
above, but with some crucial differences:

\item{(1.)} We cannot shift $\phi$ and therefore all the masses are
physical.

\item{(2.)} The one-loop computation is again divergent and we
absorbed an additive (non-universal) divergent renormalization in $t_0$.

\item{(3.)} There could again be finite corrections to $t_0$.  On
dimensional grounds they have to be linear in the masses $m_i$.  For
$N_f \ge 2$ the global $SO(2N_f)$ symmetry of the theory with $m_i=0$
prevents such corrections.

\item{(4.)} The expression \tsutwo\ makes sense only when $t(\phi)$ is
non-negative.  This means that $t_0$ should be non-negative and $m_i$
are constrained.  For $N_f> 8$ $\phi$ is bounded.  As in the $U(1)$
theories this reflects the lack of renormalizability of the theory
signaling that it must be embedded in another theory at short distance.

\item{(5.)} For $N_f=8$ and  $t_0>2\sum |m_i|$ there is no singularity in
the moduli space.  If we try to set all $m_i=0$ and then take the strong
coupling limit $t_0 \rightarrow 0$, the metric on the entire moduli
space vanishes and hence the limit is meaningless.

\item{(6.)} The theories with $N_f\le 7$ are very interesting and will be
discussed below.

The physical particles at the singularities of \tsutwo\ with non-zero
$\phi$ consist of one of the elementary gauge bosons and one of the
elementary quarks.  The low energy theory is thus the $A_0$ theory
discussed above.  (Note that the singularity in $t$ is exactly as in
that theory).  If $k$ masses are equal but do not vanish, there is a
singularity with $k$ massless charged particles.  The theory at that
singularity is the $A_{k-1}$ theory,

New theories are found around $\phi=0$.  If $N_f$ quark masses vanish,
the global symmetry is $SO(2N_f)$ and we will refer to the theory as
$D_{N_f}$. There is also another interesting $U(1)$ symmetry whose
conserved charge is the instanton number and whose conserved current is
$j={}^*(F\wedge F)$.  It does not act on the massless particles.  Here,
the massless particles at the singularity are the three gauge bosons and
$N_f$ massless quark hypermultiplets.  Also, there is a Higgs branch
isomorphic to the moduli space of $SO(2N_f)$ instantons emanating {}from
that point.

For $N_f \le 7$ we can consider the strong coupling limit $t_0
\rightarrow 0$.  The main point of \fived\ was that this limit exists.
The theory at the singularity at $\phi=0$ is a non-trivial interacting
fixed point of the renormalization group.  The notion of a particle at
this point is ill defined.  For $m_i=0$ these theories haves $E_{N_f+1}$
global symmetry with $E_5=Spin(10)$, $E_4=SU(5)$, $E_3=SU(3)\times
SU(2)$, $E_2=SU(2)\times U(1)$ and $E_1=SU(2)$.  Clearly, their Coulomb
branch is $\bR^+$.  The function $t$ is obtained {}from \tsutwo
\eqn\tenl{t(\phi)=(16-2N_f)\phi.}
Less obvious is the fact that their Higgs branch is isomorphic to the
moduli space of $E_{N_f+1}$ instantons. (For $N_f=2$ there are two
separate Higgs branches---along one of them $SU(3)$ is broken and along
the other $SU(2)$ is broken).

It is convenient to think of the parameters in these theories as
background superfields \nonren.  In these theories they are scalars in
background vector superfields \aps\ associated with the global symmetry
of the theory.  This interpretation makes it obvious that they have
dimension one. The D-term equations force them to lie in the Cartan
subalgebra.  Therefore, their number is the rank of the algebra.
Indeed, our $A_{N_f-1}$ theory has $N_f-1$ parameters ($N_f$ masses, one
linear combination of which can be removed by redefining $\phi$), and
the $D_{N_f}$ theory has $N_f$ parameters (the $N_f$ masses) while
deformations of the gauge coupling $t_0$ {}from its non-zero value is an
irrelevant parameter.  In the $E_{N_f+1}$ theory there are $N_f+1$
parameters, namely, $m_i$ and $t_0$.

Since the parameters $t_0$ and $m_i$ have dimension one, the
corresponding operators are relevant at the fixed points.  Turning on
$t_0$ with $m_i=0$ takes us to the $D_{N_f}$ theory.  Flows between
these theories are obtained by turning on some of the $m_i$.  To keep
$t=0$ at $\phi=0$ we should tune $t_0$ to $2 \sum_i |m_i|$.  Then, the
parameter
``$t_0$'' of the low energy theory is $t_0-2 \sum_i |m_i|$.

Let us consider the case $N_f=1$ in more detail.  As we said
above, for $N_f \ge 2$ the global $SO(2N_f)$ helped us identify $t_0$ as
an $SO(2N_f)$ singlet with no finite additive mass renormalization in
\tsutwo.  For $N_f=1$ the global symmetry is $E_2=SU(2) \times U(1)$.
The single mass $m_1=m$ is associated with the $U(1)$ factor.  We would
like to identify the parameter $m_0$ associated with the $SU(2)$
breaking.  The singularities of the gauge coupling occur at $t_0=2|m|$.
Since there is no symmetry under which $m$ changes its
sign\foot{The $D_{N_f}$ theories have $SO(2N_f)$ but not $O(2N_f)$
symmetry.}, the natural variable $m_0$ can be a linear combination of
$t_0$ and $m$.  Without loss of generality we define
\eqn\mzerod{m_0=t_0 -2m}
such that
\eqn\tetwo{t(\phi)=m_0 +2m +16\phi - |\phi - m| - |\phi + m|.}
This expression is valid only for $m_0+4m >0$.  Since $m_0$ is
associated with the $SU(2)$ factor, the Weyl subgroup identifies $m_0$
with $-m_0$.  To keep the equations simple we will restrict $m_0$ to be
non-negative (as with $\phi$).

For $m_0=m=0$ the theory is our $E_2$ theory.  Turning on $m_0$ with
$m=0$ leads to a flow to the trivial $D_1$ theory.  If we instead turn
on $m>0$ but keep $m_0=0$, we flow to the $E_1$ theory with $E_1=SU(2)$
global symmetry (see figure 1).  At $\phi=m$ we find an $A_0$ theory.
For non-zero $m_0$ and $m>0$ we flow to the trivial $D_0$ theory.
Again, at $\phi=m$ we find an $A_0$ theory.

Since there is no symmetry changing the sign of $m$, we do not expect
similar results for $m<0$.  In particular, there is a singularity at
$m_0+4m =0$ without a counterpart for positive $m$.  For $m_0+4m >0$ and
$m<0$ the perturbative analysis is still valid and the low energy theory
at $\phi=0$ is $D_0$ and at $\phi=-m$ it is $A_0$.  The lack of $m
\rightarrow -m$ symmetry appears only in the massive modes.  Therefore,
we will distinguish this theory {}from $D_0$ by denoting it $\tilde
D_0$.  The distinction between them becomes important only when we
approach $m_0+4m =0$.  At this point the gauge coupling diverges $t=0$.
However, the transition to this point is different {}from the transition
{}from $D_0$ to $E_1$.  The $E_1$ theory has $SU(2)$ global symmetry
while the theory at $m_0+4m =0$ has only $U(1)$ global symmetry.
Therefore, this theory must be a new non-trivial fixed point in five
dimensions.  We will refer to it as $\tilde E_1$.  Clearly it has only
one relevant parameter $s=m_0+4m$.  Using arguments similar to those in
\fived, we learn that there is no Higgs branch emanating {}from it.

\iffigs
\midinsert
$$\vbox{\centerline{\epsfxsize=4in\epsfbox{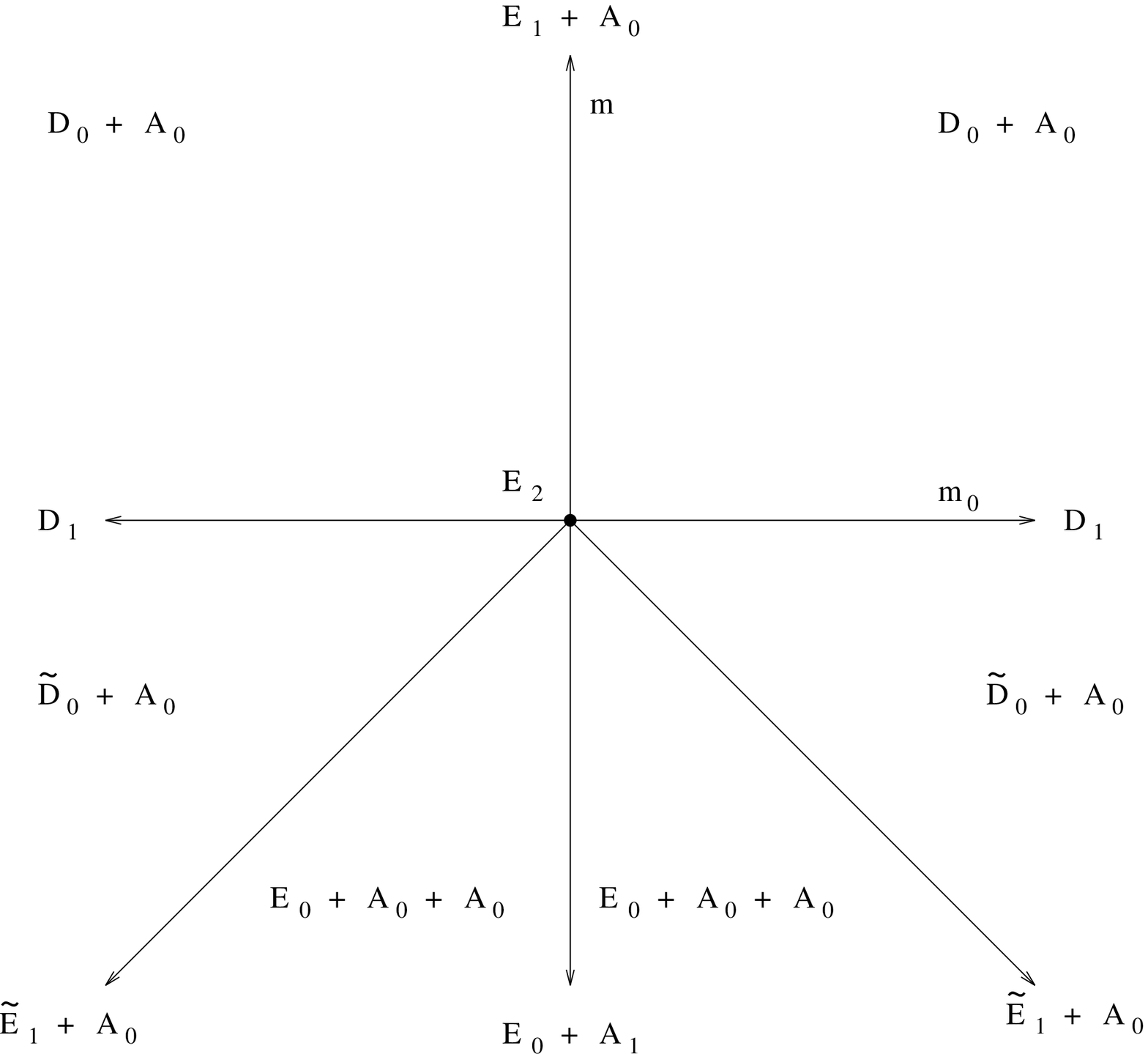}}
\centerline{Figure 1. $E_2$ and related theories.}}$$
\endinsert
\fi

We now explore the region $m_0+4m < 0$ (as before with $m_0$
non-negative). We do that by turning on the only relevant parameter in
$\tilde E_1$, namely, $s=m_0+4m$.  For $s>0 $ we flow to $\tilde D_0$.
What happens for $s<0$?  We propose that the transition is similar to
other transitions we saw before where an $A_0$ theory leaves the origin.
The theory at the origin is then at a new non-trivial fixed point with
$t=0$ at $\phi=0$ which we will call $E_0$.  Before arguing that such a
theory indeed exists, let us use this proposal to describe the physics
for $s<0$.  The global $U(1)$ symmetry of the $\tilde E_1$ theory acts
on the particle which leaves the origin at the transition.  Therefore,
there is no symmetry acting on the massless modes in $ E_0$.  This fact
is consistent with the lack of relevant parameters in this theory.

The behavior near $\phi=0$ discussed above does not affect the $A_0$
theory near $\phi=-m$.  This should remain unaffected even for $m_0+4m <
0$ and in particular at $m_0=0$ where the global $SU(2)$ symmetry must
be restored.  Now, let us track the $A_0$ theory which leaves the
origin.  It corresponds to a $U(1)$ gauge theory with one electron with
mass (at $\phi=0$) $-\alpha s$ ($\alpha $ is a positive proportionality
factor which we will determine momentarily) such that the singularity is
at $\phi = - \alpha s $.  As we make $s$ more negative (say with fixed
$m$), this singularity approaches the singularity at $\phi=-m$.  For
$m_0=0$ we must regain the global $SU(2) $ symmetry.  The only way this
could happen out of our setup with $E_0$ at the origin without
symmetries and the two $A_0$ singularities is if the $A_0$'s coalesce to
an $A_1$ theory (see figure 1).  This fixes $\alpha = {1 \over 4}$.

Assuming that $t=0$ at $\phi=0$ in the region with $E_0$ (i.e., that
the origin of $\phi$ does not move), we can find
a simple formula for the function $t$ which is valid for all $\phi \ge
0$ and $m_0 \ge 0$:
\eqn\finfo{t (\phi) = {m_0 \over 2} +3|m+ {m_0\over 4}| + 17 \phi -
|\phi+m|-|\phi-m| -|\phi +m +{m_0 \over 4}|.}
It exhibits all the singularities we discussed above.

This phenomenon has a generalization to higher $E_n$.  The $t_0$
perturbation breaks $E_n$ to $SO(2n-2) \times U(1)$.  Turning on also
equal masses to all the quarks $m_i=m$ breaks the symmetry to $SU(n-1)
\times U(1) \times U(1)$ with an $A_{n-2}$ singularity away {}from the
origin and a $D_0$ singularity at the origin.  There is another way to
get to this point.  First, we can break $E_n$ to $SU(n) \times U(1)$ by
choosing suitable values for $t_0$ and $m_i$.
We argue that in this case we have an $E_0$
theory at the origin and an $A_{n-1}$ theory away {}from the origin.  As
the symmetry is further broken to $SU(n-1) \times U(1) \times U(1)$ the
$A_{n-1}$ singularity splits to an $A_{n-2}$ and an $A_0$ singularity.
This can move to the origin, be absorbed in it to turn it into an
$\tilde E_1$ theory which can then become $\tilde D_0$, which can then
become $D_0$ exactly as in the other way of getting there.

As in \fived, all this has a simple translation to the physics of branes
in string compactification.  We use a D4-brane probe in the
compactification of the type I$'$ theory on $\bS^1/\IZ_2$.  This theory
is dual to the heterotic string compactified on $\bS^1$.  The $A_n$
theories are obtained when $n+1$ background D8-branes coalesce in the
interior of $\bS^1/\IZ_2$.  The $D_n$ theories are obtained when $n$
D8-branes are at an orientifold at the boundary of $\bS^1/\IZ_2$.  The
$E_n$ theories are obtained when the string coupling diverges on the
boundary.  The existence of the new $E_0$ theory implies that we can
have more than 16 background D8-branes in the interior.  In particular,
if the theory at one of the boundaries is at $E_0$, we can have 17
background D8-branes and if the two boundaries are at $E_0$, we can have
18 background D8-branes.  If these 18 D8-branes coalesce, the symmetry
group of the string theory is $SU(18) \times U(1)$.  In terms of the
dual heterotic string the $SU(18)$ factor comes {}from the left movers
and the $U(1)$ {}from the right movers.\foot{Denote by $\bf n$ the
representation of $SU(18)$ Kac--Moody level one whose highest weight
vector is in a representation of $SU(18)$ constructed as an
antisymmetric product of $n$ fundamentals.  Denote its character by
$\chi_{\bf n}(q)$.  Then, the partition function $|\chi_{\bf 0}
+\chi_{\bf 6} + \chi_{\bf 12}|^2 + |\chi_{\bf 3} +\chi_{\bf 9} +
\chi_{\bf 15}|^2 $ is an exceptional modular invariant of a consistent
conformal field theory.  Denote by $\psi_{\bf 0}$ and $\psi_{\bf 1}$ the
characters of the two representations of $SU(2)$ Kac--Moody level one.
Then the partition function $Z=(\chi_{\bf 0} +\chi_{\bf 6} + \chi_{\bf
12}) \overline {\psi_{\bf 0}} + (\chi_{\bf 3} +\chi_{\bf 9} + \chi_{\bf
15}) \overline \psi_{\bf 1} $ is modular invariant.  This is the
partition function of the conformal field theory based on the $(17,1)$
signature lattice for this $SU(18)$ theory.} Note that although the
string coupling diverges at the two boundaries, it can be arbitrarily
weak at the point in the middle of $\bS^1/\IZ_2$ where the background
D8-branes are.  Similarly, we could have one boundary with $E_0$ and 17
background D8-branes near an orientifold at the other boundary yielding
an $SO(34) \times U(1)$ symmetry.  Again, the physics near the
orientifold where the $SO(34)$ gauge bosons are can be arbitrarily
weakly coupled.

Finally, we argue for the existence of the $E_0$ theory.  The presence
of $SU(18)\times U(1)$ and $SO(34) \times U(1)$ points in the moduli
space of the heterotic string and the symmetry-breaking pattern tell us
that the type I$'$ theory must have corresponding points where 18
D8-branes coincide, or 17 D8-branes are at an orientifold.  This also
lets us reproduce the transitions to the $\tilde E_1$ and $\tilde D_0$
theories, guaranteeing that our scenario for the phase diagram and the
existence of the $E_0$ point is correct.

\newsec{Extremal transitions in the geometry}

In the previous section, we have described supersymmetric
five-dimensional field theories with a one-dimensional Coulomb branch,
and given the D4-brane interpretation of these theories.  We now turn to
the other string-theoretic application of the field theory analysis:
M-theory compactified on a Calabi--Yau threefold.

In such M-theory compactifications, the expectation values for the
scalars in vector multiplets are parameterized by the K\"ahler classes
of volume one on the Calabi--Yau threefold \refs{\ccdf, \fkm, \fms}.
The K\"ahler cone (i.e., the set of all K\"ahler classes) has various
boundary ``walls,'' and we should expect interesting physics as we
approach such a wall since the metric on the Calabi--Yau threefold is
degenerating there.  Walls along which $\int J\wedge J\wedge J$ vanishes
will be at infinite distance {}from the interior of the moduli space,
since we must divide a K\"ahler class $J$ by $\root3\of{\int J\wedge
J\wedge J}$ in order to get a class of volume one.\foot{This is the
analogue in five dimensions of Hayakawa's criterion \hayakawa\ in four
dimensions, which specifies which singularities are at finite distance
in the Zamolodchikov metric.} Any wall along which $\int J\wedge J\wedge
J$ does not vanish corresponds to the collapse of some proper subset of
the Calabi--Yau threefold \wilson, for if the entire Calabi--Yau
collapses, then $\int J\wedge J\wedge J$ must vanish.

The possible ways that a proper subset of a Calabi--Yau threefold can
collapse upon approaching a boundary wall of the K\"ahler cone have been
analyzed in considerable detail in the mathematics literature.  We
restrict our attention to codimension-one walls, so that approaching the
wall involves tuning only a single vector multiplet expectation value;
we also assume that the complex structure on the Calabi--Yau threefold
is generic (so that no hypermultiplets need to be tuned to reach the
singularity).  Under these assumptions, the possible types of
``collapse'' are the following:\foot{When the complex structure on the
Calabi--Yau threefold is not generic, there are further types of
collapse which can occur, described in the appendix.  We should point
out here that the possibility is left open by the literature that one or
two additional cases might need to be included along with the ones we
present.}

\item{(1.)} A collection of $N$ disjoint $\Bbb{P}^1$'s can collapse to
$N$  ``ordinary double points.''

\item{(2.)} A rational surface which has a ``birational ruling,'' $N$
fibers of which consist of a pair of $\Bbb{P}^1$'s and the remaining
fibers of which are irreducible $\Bbb{P}^1$'s, can collapse
along the ruling to a $\Bbb{P}^1$
of singularities.

\item{(3.)} A ``del Pezzo surface''\foot{A {\it del Pezzo surface} is a
complex manifold $S$ of complex dimension $2$ whose anticanonical
divisor $-K_S$ is ample, i.e., $-K_S\cdot C>0$ for every Riemann surface
$C$ contained in $S$.  (See \refs{\demazure,\manin}, or for a brief
account in the physics literature, \MVii.)  Every such surface is known
to be either $\Bbb{P}^1\times \Bbb{P}^1$, or a blowup of $\Bbb{P}^2$;
when more than one point of $\Bbb{P}^2$ has been blown up the surface
can also be described as a blowup of $\Bbb{P}^1\times\Bbb{P}^1$.} which
is the blowup of $\Bbb{P}^1\times \Bbb{P}^1$ at $N$ points ($0\le
N\le7$), or of $\Bbb{P}^2$ at $N+1$ points ($0\le N+1\le8$), can
collapse to a point.  (When $N\ge1$, the two descriptions are
equivalent.)

\noindent
Anticipating our field theory identifications, we shall refer to these
three cases as $A_{N-1}$, $D_N$, and $E_{N+1}$ (or $\tilde E_1$)
respectively (where $E_1$ is the case of $\Bbb{P}^1\times \Bbb{P}^1$
blown up at $0$ points, $\tilde E_1$ is the case of $\Bbb{P}^2$ blown up
at $1$ point, and $E_0$ is the case of $\Bbb{P}^2$ blown up at $0$
points).

Although the ``collapse'' in each of these cases is the result of a
degeneration of the metric on the Calabi--Yau, the resulting collapsed
space can be described using algebraic geometry as the solution set of a
collection of polynomial equations.  (The algebraic versions of such
collapses are known as ``extremal contractions'' \mori, and have been
studied extensively in the mathematics literature.)  Since the
singularities have an algebraic description, algebraic methods can be
used to determine whether or not the collapsed space can be smoothed out
by varying the coefficients in the polynomial equations.  If this is
possible, the smoothed space is again a Calabi--Yau threefold, but of a
completely different topology than the original one.  The process of
collapsing via an extremal contraction and then smoothing out the
resulting space to produce a new Calabi--Yau threefold is known as an
``extremal transition'' \look.

It will be useful to consider a slight generalization of this setup, in
which the geometry of the contraction is similar but the contraction may
occur at higher codimension in the K\"ahler cone.  Let $S$ be either (1)
a collection of $\Bbb{P}^1$'s, (2) a birationally ruled surface over
$\Bbb{P}^1$, or (3) a del Pezzo surface, and suppose that $S$ is
embedded in a Calabi--Yau threefold $X$ in such a way that there is a
contraction $X\to\overline{X}$ associated to some wall of the K\"ahler
cone of $X$ which maps $S$ to (1) ordinary double points, (2) a
$\Bbb{P}^1$ of singularities, or (3) a singular point, respectively.  If
the codimension of the wall of the K\"ahler cone is $k$, then the
dimension of the image of the natural map $H^{1,1}(X)\to H^{1,1}(S)$ is
$k+\epsilon$, where $\epsilon=0$ in cases (1) and (3), and $\epsilon=1$
in case (2).

In the $A_{N-1}$ case (in this more general sense),   a physical
interpretation of the contraction and extremal transition was found in
\refs{\rStrominger,\rGMS} when
the type IIA string is compactified on the Calabi--Yau
threefold:\foot{The analysis of \refs{\rStrominger,\rGMS} was done in
type IIB language, but as has been pointed out in \refs{\rBBS,\rGMV},
the same analysis can be applied to the type IIA string.} there are $k$
$U(1)$'s associated to the homology classes spanned by the
$\Bbb{P}^1$'s, there are $N$ massive hypermultiplets charged under
$U(1)^k$ which become simultaneously massless at the transition point,
and there are $N-k$ flat directions at the transition which produce
$N-k$ new parameters (the Higgs branch in the field theory).
Mathematically, it was already known that this extremal transition
exists \friedman, and that the number of new complex deformation
parameters would be $N-k$ \refs{\clemens,\schoen}.  (These parameters,
together with their Ramond-Ramond partners, will give the Higgs branch
of the field theory.)  The five-dimensional version of this transition
was discussed in \mandf, where it was seen that the field theory
description of $A_{N-1}$ type as described in the previous section
applies.

The $D_N$ theories are obtained when a birationally ruled surface over a
$\Bbb{P}^1$ base collapses.  The relevant light particles arise by
wrapping a two-brane over various two-cycles.  The quantization of
their collective coordinates proceeds as in \mandf\ and determines their
quantum numbers.  Wrapping the two-brane over the entire collapsing
fiber we find vector multiplets in the adjoint representation of
$SU(2)$.  In the case analyzed in \mandf\ the base was a Riemann surface
of genus $g$ and hence there were also $g$ hypermultiplets in the
adjoint representation which we do not have.  Instead, we get
hypermultiplets from the 2$N$ special $\Bbb{P}^1$'s in our fiber.  As in
the $A_k$ cases, every such $\Bbb{P}^1$ leads to one hypermultiplet
which is charged under the unbroken $U(1)$ with the minimal unit of
charge.  Therefore, every pair of $\Bbb{P}^1$'s leads to a
hypermultiplet which is an $SU(2)$ doublet.  To summarize, the low
energy spectrum is that of an $SU(2)$ gauge theory with $N$
hypermultiplets which are $SU(2)$ doublets.  This is the spectrum of the
$D_N$ field theories of the previous section.

When the surface $S$ collapses, there is an effective field theory
description near $\phi=0$ (which labels the point of collapse
in the K\"ahler moduli space).  The gauge coupling will take
the form $t(\phi)=t_0+c\phi$; since $c$ also multiplies the Chern--Simons
term in the action, it can be calculated in terms of the intersection
theory on $X$ \refs{\ccdf, \fkm, \fms} and it turns out
to be $c=2S\cdot S\cdot S$.  (The factor of $2$ is a normalization designed
to match our field theory conventions.)
More generally, the second derivative of the prepotential with
respect to the field corresponding to $S$ takes the form $2S\cdot S\cdot H$,
where $H$ is an arbitrary divisor from the K\"ahler cone on $X$.
Expanding $H$ in terms of a basis of $H^{1,1}(X)$ which contains
$S$, we obtain an expression for the gauge coupling of the
form
\eqn\ecoupmain{\eqalign{
t(\phi)&=2S\cdot S\cdot (\sum\alpha_iH_i+\phi S)\cr
&=2S\cdot S\cdot(\sum\alpha_iH_i)+c\phi=t_0+c\phi\ .
}}
Of course, any divisors $H_i$ with $S\cdot S\cdot H_i=0$ can be
omitted from this expression.

In the simplest version of the $D_N$ theories---the ones with $k=1$---there
are only two independent elements in $H^{1,1}(X)$ which are not
orthogonal to $S\cdot S$; we choose a basis for these consisting of $S$ and a
divisor $H_0$ with $H_0\cdot S={1\over4}{\cal F}$, where ${\cal F}$ is a
generic fiber of the ruling.  Then the gauge coupling is given by
\eqn\efieldone{t(\phi)=2\,S\cdot S\cdot(t_0 H_0+\phi S) =
-2K_S\cdot({t_0\over4}{\cal F}-\phi K_S)= t_0 + (16-2N)\phi\ .}
The factor of $1/4$ was included in the definition of $H_0$ to make
the constant term match our conventions from section 2.

At the other extreme, if $k=N+1$, we will find that all of the
parameters in the field theory can be realized in the geometry.  In
addition to the divisors $H_0$ and $S$, we choose divisors $H_i$ on $X$
such that $ H_i\cdot S={1\over2}{\cal F}-\Gamma_i$, where $\Gamma_i$ is
one of the components of a reducible fiber.  Note that the other
component $\Gamma_i'$ of that fiber then satisfies $(-H_i)\cdot
S={1\over2}{\cal F}-\Gamma_i'$, so the choice of component only affects
the sign of $H_i$.\foot{However, as we learned in section 2, the sign is
important; the one given here will be the correct choice once we get to
the $E$ theories.} If we write a general divisor in the form $t_0
H_0+\sum_{i=1}^N m_i H_i +\phi S$ and assume it lies in the K\"ahler
cone of $X$, then the gauge coupling is given by
\eqn\efieldtwo{
t(\phi)=2\,S\cdot S\cdot(t_0 H_0 + \sum_{i=1}^N m_iH_i + \phi S) =
t_0+(16-2N)\phi \ ,}
since $S\cdot S\cdot H_i=0$.
The K\"ahler cone of $X$ is determined in part by the conditions
\eqn\ebounds{\eqalign{
0<(t_0 H_0+\sum_{i=1}^N m_i H_i +\phi S)\cdot {\cal F}&=2\phi\cr
0<(t_0 H_0+\sum_{i=1}^N m_i H_i +\phi S)\cdot \Gamma_i&=\phi+m_i\cr
0<(t_0 H_0+\sum_{i=1}^N m_i H_i +\phi S)\cdot \Gamma_i'&=\phi-m_i \ .
}}

The curves $\Gamma_i$ or $\Gamma_i'$ can now be flopped to produce other
birational models of $X$.  Pick disjoint subsets $A$ and $B$ {}from
$\{1,\dots,N\}$, and flop $\Gamma_i$ when $i\in A$ and $\Gamma_i'$ when
$i\in B$.  Then the K\"ahler cone of the flopped model is determined
in part by the conditions
\eqn\emorebounds{\eqalign{
0<2\phi ,& \cr
0<\phi+m_i ,&\ \hbox{for}\ i\not\in A\cr
0<-\phi-m_i ,&\ \hbox{for}\ i\in A\cr
0<\phi-m_i ,&\ \hbox{for}\ i\not\in B\cr
0<-\phi+m_i ,&\ \hbox{for}\ i\in B\ .
}}
There is also a new expression for the gauge coupling
valid in that cone:
\eqn\efieldthree{
t(\phi)=t_0+(16-2N+2\#(A)+2\#(B))\phi+2\sum_{i\in A}m_i-2\sum_{i\in B}m_i}
(since $S\cdot S\cdot H_i$ decreases by $1$ for $i\in A$ and increases
by $1$ for $i\in B$ when we move to the new cone).  These formulas can
all be put together into a single piecewise linear function
\eqn\ePL{t(\phi)=2S\cdot S\cdot(t_0 H_0+\sum_{i=1}^Nm_iH_i+\phi S)
=t_0 +16\phi -\sum_{i=1}^N|\phi+m_i|-\sum_{i=1}^N|\phi-m_i|\ ,}
valid throughout the union of all of the K\"ahler cones, and precisely
reproducing eqn.~\tsutwo.  This firmly establishes our identification of
not only the field theory, but also the parameters in it.

Now we have enough data to identify the interacting field theories
$E_{N+1}$ in geometric terms.  We start with the $D_N$ theory, realized
geometrically as above with some value of $k$.  Notice that the surface
$S$ in this case is actually a blowup of $\Bbb{P}^1\times \Bbb{P}^1$ at
$N$ points.  We assume $N\le7$, and let $t_0\to 0$ and $\phi\to0$, and
well as $m_i\to0$ for any parameters $m_i$ which are present.  This
means that we are approaching a part of the cone in which the entire
surface $S$ collapses to a point, i.e., $X$ experiences the collapse of
a del Pezzo surface.  In other words, {\it the del Pezzo contractions
precisely realize the interacting five-dimensional field theories}. (We
have not yet established this in the cases of $E_0$ and $\tilde E_1$,
but we shall do so shortly.)

As in the $A_N$ and $D_N$ cases we can examine the BPS states which
approach zero mass at the transitions.  First, we could wrap various
two-cycles with two-branes to find light particles.  We could also wrap a
five-brane over the entire collapsing surface $S$ to yield a string
whose tension approaches zero.  This string is electric-magnetic dual to
the light particles.  Such a spectrum in five dimensions was first
observed in \mandf\ for one of the del Pezzo contractions.  Having both
massless particles and tensionless strings at the transition points
shows that they are not free field theories.  We identify them with
interacting field theories.

Before exploring the geometry of these del Pezzo contractions further,
we pause to point out one of the striking consequences of this
identification: {\it the Higgs branches of generic extremal transitions
are given by an $A$, $D$, or $E$ instanton moduli space}.  This
prediction serves to unify a collection of disparate known facts about
these transitions, which had not previously been seen as part of any
pattern.  Let us review what is known mathematically.

Given $S\subset X$ contracted by $X\to\overline{X}$, the mathematical
question is whether $\overline{X}$ can be smoothed
by varying its equations, and if so, what is the number of new parameters,
i.e., the difference $\dim \Def(\overline{X})-\dim \Def(X)$ between the
dimensions of the deformation spaces.  (Note that this difference will coincide
with the dimension of the Higgs branch in the field theory, although
strictly speaking before making such a comparison we should construct
the hypermultiplet moduli space in the string theory which requires the
inclusion
of Ramond-Ramond partners to the complex structure moduli.)  The
known facts about this mathematical question are as follows:
\item{(1.)} In the $A_{N-1}$ case, as pointed out above, an extremal
transition exists whenever $N>k$ \friedman\ and the number of new
parameters is $N-k$ (as is seen by calculating $h^{2,1}$
\refs{\clemens,\schoen}).
\item{(2.)} In the $D_N$ case, assuming for simplicity that
$k=1$,\foot{This is the assumption ``$X$ is $\Bbb{Q}$-factorial'' which
sometimes appears in the mathematics literature.} an extremal transition
exists whenever $N\ge2$ \grossprim.  The case $N=2$ is somewhat special,
in that the deformation space for $\overline{X}$ is not smooth, but has
two
components of dimension one \grossobstr\foot{This was only shown in an
example, but it probably holds in general.} (as expected {}from the
field theory, since $D_2$ is not simple).  Our prediction for the number
of new parameters is $2N-k-2$ in general,
and this is consistent with all known
facts, including the examples worked out in \klkam.
\item{(3.)} The $E_{N+1}$ cases are rather complicated, and one must
consider different values of $N$ separately (cf.~\grossdef); we again
assume $k=1$ for simplicity.
\itemitem{(a.)} For $\tilde E_1$, and $E_{N+1}$, $0\le N+1\le 3$, the
singularity of $\overline{X}$ is toric and the deformation theory has
been worked out by Altmann \altmann; the result is that there is no
smoothing for $E_0$ or $\tilde E_1$ (this is a classic result of
Schlessinger \quotient\ in the $E_0$ case, which is a $\Bbb{Z}_3$
quotient singularity), a one-dimensional deformation space for the cases
of $E_1$ and $E_2$, and a deformation space with two components, one of
dimension one and one of dimension two, in the case of $E_3$.
In addition, there are further, obstructed first-order deformations in
precisely
the $\tilde E_1$ and $E_2$ cases---the ones with a $U(1)$ factor in
their symmetry groups.
\itemitem{(b.)} For $E_4$, the singularity of $\overline{X}$ is defined
by a Pfaffian ideal and the theory of those can be used to study the
deformations \refs{\BE,\kleplak}; the result is that a smoothing exists.
\itemitem{(c.)} For $E_{N+1}$, $5\le N+1\le 8$, the singular of
$\overline{X}$ is a complete intersection and the deformation theory is
relatively easy; smoothings exist in all cases.
\item{\hphantom{(1.)}} For these del Pezzo contractions, it was observed in
\MVii\ that at least for $N+1\ge5$, the number of new parameters for
this extremal transition should be $c_{N+1}-k$, where $c_{N+1}$ is the
Coxeter number of the group $E_{N+1}$. In fact, this statement should
continue to hold for all values of $N$, if properly interpreted: when
the group $E_{N+1}$ is not simple, there must be two components whose
dimensions are given by the Coxeter numbers (minus $k$) of the two
factors in the group.  (One should also include an obstructed deformation
if there is a $U(1)$ factor.) This formula is compatible with all known
calculations of dimensions of smoothing components.\foot{We thank Mark
Gross for discussions on this point.}

It is quite remarkable how this prediction about the Higgs branch
unifies phenomena such as the two components or other obstructed deformations
which appeared in sporadic
locations (and looked in the past like counterexamples to any regularity
of behavior).  It also makes clear why the main theorems of \grossdef\
and \grossprim\ are so similar (as remarked in \grossprim).

As in the $D_N$ cases, when $k>1$ so that some of the field theory
parameters are realized in the string theory, the $E_{N+1}$ theories
will have a rich structure with various possible flops leading to new
(K\"ahler) cones in which the gauge coupling takes a different form.
This structure can be completely analyzed using the known combinatorial
structure of the del Pezzo surfaces.  In brief, the condition which must
be satisfied in order to flop some collection of rational curves is that
the del Pezzo surface can be blown down along those curves.  The Weyl
group of $E_{N+1}$ acts in a natural way\foot{This was known a long time
ago \refs{\demazure,\manin}, and provides an important clue to the
$E_{N+1}$ symmetry in the physics \MVii.} on $H^{1,1}(S)$, and permutes
the various collections of curves which can be blown down.  This can be
used to write down the complete phase diagram for each case, and of
course the gauge coupling can be directly calculated in every phase.
This is essentially the same combinatorial data which we used in the
field theory, so it is not surprising that the results turn out to be the same.

We will carry this out in the case of $N=1$, in order to complete our
geometric identifications of the interacting field theories by including
the $\tilde E_1$ and $E_0$ cases.

We take a del Pezzo surface $S$ with $N=1$, embedded on a Calabi--Yau
manifold $X$ with $k=3$.  The
surface $S$ can be regarded as a blowup of $\Bbb{P}^2$ at two points
$P_1$, $P_2$; we let $\Gamma_1$ and $\Gamma_2$ be the exceptional divisors
of the blowup.  There is another exceptional divisor on $S$, a curve
$\Gamma_3$ which is the proper transform of the line passing through
$P_1$ and $P_2$.

We also introduce three divisors $H_\ell$, $H_1$, and $H_2$, which
intersect $S$ in the classes of a line, and the two rulings.
Specifically, $H_1\cdot S=\Gamma_2+\Gamma_3$ is in the same class
as the proper transform
of lines passing through $P_1$, and $H_2\cdot S=\Gamma_1+\Gamma_3$
is in the same class as the proper transform of lines passing through $P_2$.
We write a general divisor on $X$ (modulo those orthogonal to $S\cdot S$)
 in the form
\eqn\egendiv{a_1H_1+a_2H_2+a_3H_\ell\ .}

It is known \mori\ that on any del Pezzo surface, the walls of the
K\"ahler cone are dual to the possible exceptional curves on the
surface, which in our example are precisely $\Gamma_1$, $\Gamma_2$, and
$\Gamma_3$.  Thus, the K\"ahler cone of $X$ is defined by
\eqn\econeone{\eqalign{
0<(a_1H_1+a_2H_2+a_3H_\ell)\cdot \Gamma_1 &= a_1 \cr
0<(a_1H_1+a_2H_2+a_3H_\ell)\cdot \Gamma_2 &= a_2 \cr
0<(a_1H_1+a_2H_2+a_3H_\ell)\cdot \Gamma_3 &= a_3 \ .
}}

We now consider the birational models of $X$ obtained by flopping these
curves.  We label the model by the type of surface $S$ has become on
that model, so the original Calabi--Yau is labeled $X_{Bl_2\Bbb{P}^2}$
(a two-point blowup of $\Bbb{P}^2$).  We could flop $\Gamma_3$ to get a
model $X_{\Bbb{F}_0}$ with $S=\Bbb{P}^1\times\Bbb{P}^1$.  (We introduce
the slightly simpler (standard) notation of $\Bbb{F}_0$ to denote
$\Bbb{P}^1\times\Bbb{P}^1$ and $\Bbb{F}_1$ to denote $Bl_1\Bbb{P}^2$; we
will distinguish the blowups of $\Bbb{P}^2$ at $P_1$ and $P_2$ by
labeling the latter as $\tilde{\Bbb{F}}_1$.)  On this model, $\Gamma_1$
and $\Gamma_2$ meet $H_\ell$, and the
K\"ahler cone is defined by
\eqn\econetwo{a_1+a_3>0,\quad
a_2+a_3>0,\quad -a_3>0\ .}
Or, we could flop $\Gamma_1$ to get a model $X_{\Bbb{F}_1}$ on which
$\Gamma_3$ meets $H_1$ so that the K\"ahler cone is
\eqn\econethree{-a_1>0,\quad a_2>0,\quad
a_1+a_3>0\ .}
Similarly, if we flop $\Gamma_2$ then $\Gamma_3$ meets $H_2$
and we get a model $X_{\tilde{\Bbb{F}}_1}$
whose K\"ahler cone is
\eqn\econefour{a_1>0,\quad -a_2>0,\quad
a_2+a_3>0\ .}
 Finally, we could flop both $\Gamma_1$ and $\Gamma_2$ to get
a model $X_{\Bbb{P}^2}$ with $\Gamma_3$ meeting both $H_1$ and
$H_2$ so that the K\"ahler cone is
\eqn\econefive{-a_1>0,\quad -a_2>0,\quad a_1+a_2+a_3>0\ .}
All five of these K\"ahler cones are illustrated in figure 2.

\iffigs
\midinsert
$$\vbox{\centerline{\epsfxsize=3in\epsfbox{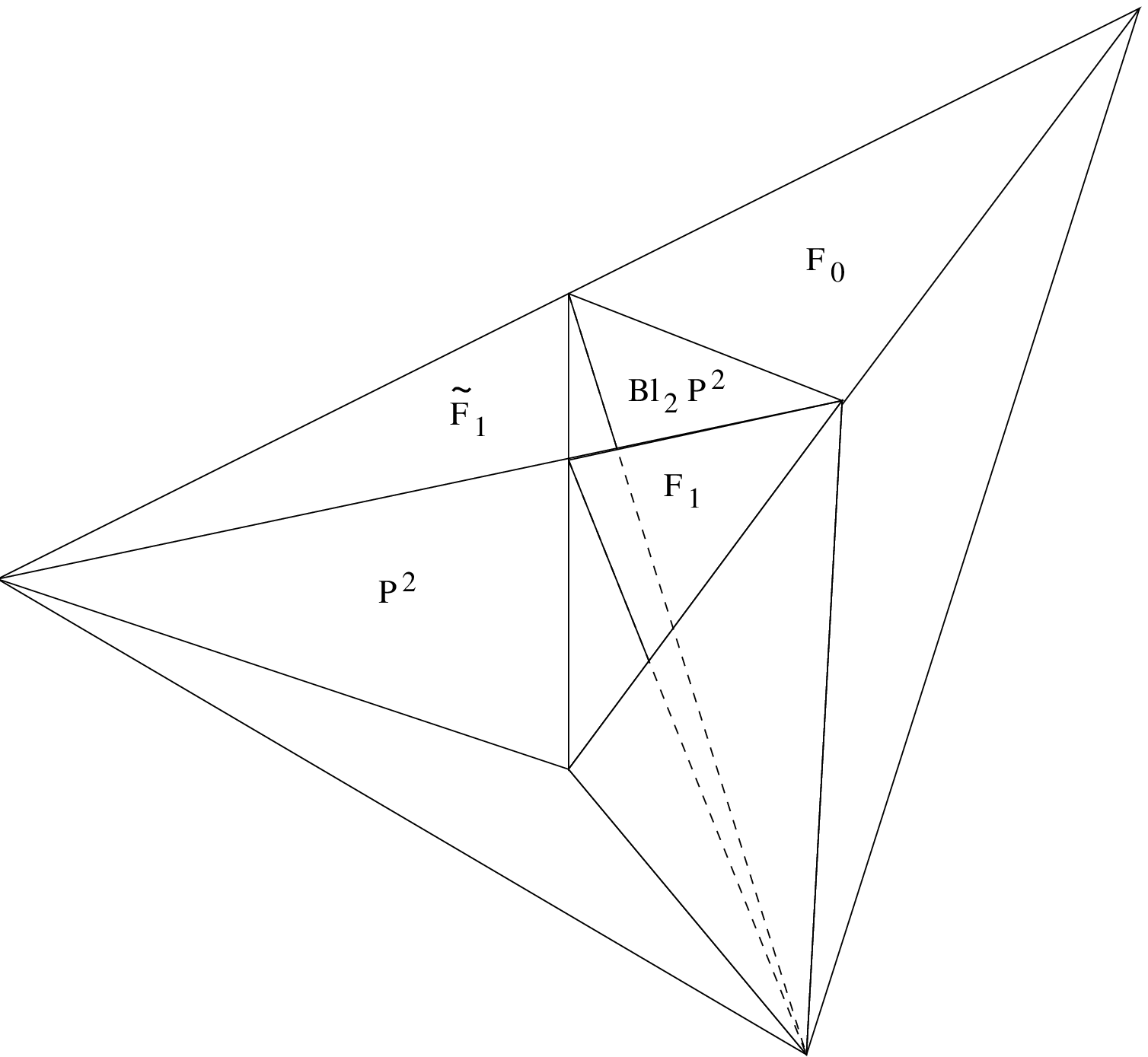}}
\centerline{Figure 2. Phases for a del Pezzo contraction with $N=1$.}}$$
\endinsert
\fi

We now ask what happens when we approach one of the boundary walls which
does not correspond to a flop.  We must either have a ruling along which
the surface is contracted, or the surface must be a del Pezzo,
contracted by its generator.  The $\Bbb{F}_0$ case has two rulings
$\Gamma_1$ and $\Gamma_2$, and
the corresponding coefficients $a_1+a_3$
and $a_2+a_3$ when
sent to zero produce $D_0$ theories.  One sees an $E_1$ theory along the
line of intersection of those boundary walls, where the entire
$\Bbb{F}_0$ is contracted to a point.

Each of the $\Bbb{F}_1$'s has a ruling (given by $\Gamma_3$),
 and contraction of these
produces further $D_0$-type theories.  To facilitate comparison with the
field theory analysis, we label these particular theories as
``$\tilde{D}_0$ theories,'' although as in the field theory case the
differences are minor.  At the boundary between the contraction of one
of the $\Bbb{F}_1$'s and the contraction of the corresponding ruling on
$\Bbb{F}_0$ (as detected by the coefficients being the same),
we find a $D_1$ theory---the entire surface
$Bl_2\,\Bbb{P}^2$ is contracted along the ruling in this case.

Finally, along the boundary of the $\Bbb{P}^2$ cone at which
$a_1+a_2+a_3$ goes to zero, we get an $E_0$ theory.  Between
this $E_0$ theory and the neighboring $\tilde{D}_0$ theory one finds an
$\tilde{E}_1$ theory, in which the $\Bbb{F}_1$ or $\tilde{\Bbb{F}}_1$
has been contracted to a point.  Furthermore, at the vertex of this part
of the cone, all three classes {}from $Bl_2\,\Bbb{P}^2$ have been
contracted and we get a theory of type $E_2$.

These boundary behaviors are illustrated in figure 3, where to simplify
the drawing we show only a slice of the cone {}from figure 2 (so the
$E_2$ point is omitted entirely).  To compare with the field theory
illustrated in figure 1, one should imagine taking the boundary of the
cone of figure 2 and flattening it onto the plane.  The labels along the
edges in figure 3 would then label the regions in the boundary of the
cone, which are easily seen to precisely correspond to figure 1.

\iffigs
\midinsert
$$\vbox{\centerline{\epsfxsize=3in\epsfbox{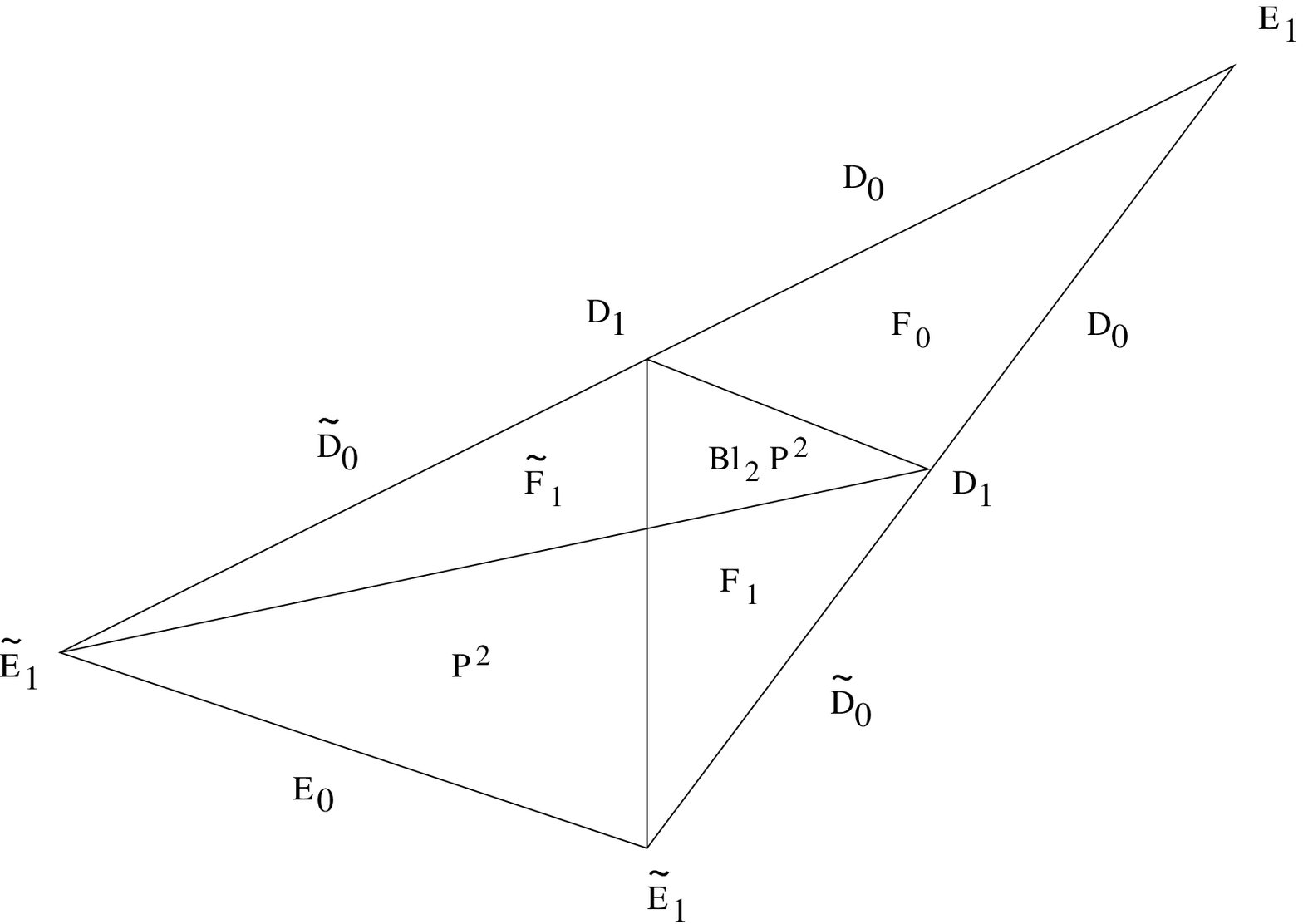}}
\centerline{Figure 3. Cross-section of the phase diagram, showing boundary
behavior.}}$$
\endinsert
\fi

In fact, it is this detailed correspondence which allows us to identify
with confidence the $E_1$ and $\tilde{E}_1$ field theories with the
$\Bbb{F}_0=\Bbb{P}^1\times\Bbb{P}^1$
 and $\Bbb{F}_1=Bl_P\Bbb{P}^2$ del Pezzo surfaces, respectively.
Further evidence that these identifications are correct is provided by
the perfect correspondence between the mathematical deformation theory
of $\overline{X}$, and the structure of the Higgs branch of the field
theory (which is different in the two cases).

\bigbreak\bigskip\bigskip\centerline{\bf Acknowledgments}\nobreak
The work of D.R.M. was supported in part by NSF grant DMS-9401447, and
that of N.S. was supported in part by DOE grant DE-FG02-96ER40559.  We
thank M. Douglas, B. Greene, M. Gross, S. Shenker, C. Vafa, N. Warner,
and E. Witten for helpful discussions.

\appendix{A}{Non-generic primitive contractions}

The list we have given in section 3 of possible primitive extremal contractions
{}from a Calabi--Yau threefold is not complete.
(A ``primitive'' contraction is one which occurs at codimension one in
the K\"ahler cone.)

In the case of curves collapsing to points, in principle any terminal
singularity could occur.  However, this should not happen at generic
complex moduli, in light of some results of Namikawa \namikawa.  Although
he did not state things precisely in this way, the proof of Theorem
(2.5) of \namikawa\ suggests that after a primitive extremal contraction
$X\to\overline{X}$ where $\overline{X}$ has terminal singularities, one
should be able to
deform away the non-ordinary double points on $\overline{X}$ while
keeping the ordinary double points. The implication would be that at generic
complex moduli of $X$ only ordinary double points should appear in such
contractions.  This question deserves further study.

In the case of a divisor collapsing to a point, the original
classification goes back to Reid's first paper on canonical
singularities \reidcanon, in which it was shown that the divisor must be
a ``generalized del Pezzo surface.''  Such surfaces include nonsingular
del Pezzo surfaces (blowups of $\Bbb{P}^2$ or $\Bbb{P}^1\times
\Bbb{P}^1$ with ample anti-canonical bundle), del Pezzo surfaces with
rational double points (which will deform to the nonsingular ones),
nonnormal del Pezzo surfaces, and cones over elliptic normal curves.  As
Mark Gross has pointed out
\grossdef, the latter can only occur as exceptional divisors of
primitive contractions {}from a nonsingular Calabi--Yau if the degree is
$1$, $2$, or $3$, and those cones will deform to del Pezzo surfaces of
types $E_8$, $E_7$ and $E_6$, respectively.

The troublesome case is the nonnormal del Pezzo surfaces \reiddelP.
They include analogous cones over nodal or cuspidal rational curves, and
Gross's argument also limits those to the three low-degree cases (which deform
to
$E_8$, $E_7$, and $E_6$).  Of the remaining possibilities, Gross shows
that only one---a nonnormal del Pezzo of degree $7$, corresponding to
some sort of degenerate form of $E_2$---could occur.  We believe that
this will not occur at generic complex structure, but have no concrete
evidence to support this belief.

Finally, we come to the most technically difficult case of a divisor $S$
collapsing to a curve, analyzed in \grossprim.  When the curve has genus
at least one, there is a deformation of $X$ under which the divisor
ceases to be effective \refs{\wilson,\grossprim}; thus, this case does
not occur when the complex structure is generic.  (This is also expected
on physical grounds \rKMP, since the field theory description consists
of $SU(2)$ with charged matter in $g\ge1$ adjoints, and possibly some
fundamentals as well.)  In the case of contraction to a curve of genus
zero, there is the possibility that curves being contracted are
generically two $\Bbb{P}^1$'s which meet; according to \grossprim, for
generic complex structure this can only happen in the case of $S^3=7$,
i.e., some kind of degenerate form of $D_1$.  We believe that this case
too will not occur at generic complex structure, but have no concrete
evidence to support this belief either.  (Note that \grossprim, Example
1.5, shows that such surfaces can indeed occur on Calabi--Yau threefolds.)

\listrefs
\end